# Propagation dynamics of the circular Airy Gaussian vortex beams in the fractional nonlinear Schrödinger equation

Shangling He[1], Kangzhu Zhou[2], Xi Peng[3], Jialong Tu[1], Yingji He[3], and Dongmei Deng[1]*

[1]*Guangdong Provincial Key Laboratory of Nanophotonic Functional Materials and Devices, South China Normal University, Guangzhou 510631, China*

[2]*School of Electronic and Computer Engineering, Peking University, Beijing 100871, China*

[3]*School of Photoelectric Engineering, Guangdong Polytechnic Normal University, Guangzhou 510665, China*

*Corresponding author: dmdeng@263.net*

## Abstract

We have investigated the propagation dynamics of the circular Airy Gaussian vortex beams (CAGVBs) in a (2+1)-dimesional optical system discribed by fractional nonlinear Schrödinger equation (FNSE). By combining fractional diffraction with nonlinear effects, the abruptly autofocusing effect becomes weaker, the radius of the focusing beams becomes bigger and the autofocusing length will be shorter with increase of fractional diffraction Lévy index $\alpha$. It has been found that the abruptly autofocusing effect becomes weaker and the abruptly autofocusing length becomes longer if distribution factor of CAGVBs increases for fixing the Lévy index $\alpha$. The roles of the input power and the topological charge in determining the autofocusing properties are also discussed. Then, we have found the CAGVBs with outward acceleration and shown the autodefocusing properties. Finally, the off-axis CAGVBs with positive vortex pairs in the FNSE optical system have shown interesting features during propagation.

1. **Introduction**

In 1982, the concept of fractality in quantum physics was introduced by B. B. Mandelbrot [1]. Then, space fractional quantum mechanics was utilized to describe the physical phenomenon that the Brownian orbits in Feynman path integrals replacing by Lévy flights, which extends the framework of fractional quantum and statistical mechanics [2].

As we know, fractional Schrödinger equations (FSEs) have been demon strated by Laskin [3], and have attracted extensive attention in physics [4-9]. Specially, the FSE producing dual Airy beams in a spherical optical cavity, is introduced by Longhi in detail [10]. Through utilization of the FSE, the propagation of beams was investigated with different external potentials and nonlinear terms [11-13]. With a longitudinal modulation potential, the period of Rabi oscillations and the efficiency of resonant conversion were also proved to be control by the Lévy index [14]. Based on the PT symmetry in the FSE, there was an important breakthrough that the linear symmetry of band struc ture was exploited in one-dimensional case [15].

Recently, nonlinear optical phenomena have attracted widely interest in studying optical solitons based on the FNSE [16-20]. Vortex solitons carried with unit topological charge, have also been reported in FNSE with space fractional lattices. The vortex soliton can only be stable in the middle region of the transmission constant, and the stability range varies with the change of Lévy index [21]. Vortex solitons in FNSE with the cubic-quintic nonlinearity have also been proposed [22].

Airy beams propagate with ring property and in nonlinear medium have also been discussed [23-27]. For example, circular Airy Gaussian beams (CAGBs) with vortex propagating free space have aroused some research [28], and the propagation characteristics of the autofocused CAGBs with a spiral phase in the nonlinear medium has been explored [29]. Based on free space, it was also found that the split of Airy beams can be governed in the FSE with Lévy index and the periodic self-imaging effect under the role of a symmetric barrier was proposed for α = 1 [30].

To date the propagation properties of autofocusing circular Airy Gaussian vortex beams (CAGVBs) in the fractional nonlinear Schrödinger equation (FNSE) optical system have not been researched. Some questions arise naturally: Can the CAGVBs propagate stably in a FNSE optical system? What propagation properties (in particular the autodefocusing properties) do the CAGVBs possess by controlling the effect of Lévy index in a FNSE optical system?

This paper consists of the following parts. In section 2, the theoretical model was introduced for the CAGVBs in the FNSE optical system. In section 3, the propagating dynamics of abruptly autofocusing on-axis CAGVBs in the FNSE optical system is discussed. Then in section 4, we investigate the propagating dynamics of off-axis CAGVBs in the FNSE optical system. Finally, the paper is concluded by section 5.

2. **The theoretical model**

We consider the propagation of the CAGVBs in a nonlinear medium obeys the FNSE

$$i\frac{\partial u}{\partial z} - \frac{1}{2kw_0^{2-\alpha}}\left(-\frac{\partial^2}{\partial x^2} - \frac{\partial^2}{\partial y^2}\right)^{\alpha/2} u + \frac{n_2 k}{n_0}|u|^2 u = 0, \quad (1)$$

where $u$ denotes the amplitude of the optical wave, $z$ is the longitudinal propagation distance, $k = 2\pi/\lambda$ denotes the wave number of the optical wave, $\lambda$ is the wavelength of the incident light, α is the Lévy index (1<α≤2), variables x and y are the scaled transverse coordinates, and $n_0$ is the refractive index of the free space and $n_2$ is the nonlinear coefficient of the Kerr medium. The fractional-diffraction operator in Eq. (1) is defined by the known integral expression [2,4,5],

$$H_\alpha u(x,y,z) = \frac{1}{2w_0^{2-\alpha}}\left(-\frac{\partial^2}{\partial x^2}-\frac{\partial^2}{\partial y^2}\right)^{\alpha/2} u(x,y,z) + \frac{n_2 k}{n_0}|u(x,y,z)|^2 u(x,y,z)$$

$$= \frac{1}{2w_0^{2-\alpha}}\iint dk_x dk_y \left(k_x^2+k_y^2\right)^{\alpha/2} \tilde{u}(k_x,k_y,z) + \frac{n_2 k}{n_0}|u(x,y,z)|^2 u(x,y,z)$$

The model's Hamiltonian is $ik\frac{\partial u}{\partial z} = H_\alpha u,$ here Hamiltonian operator is

$$H_\alpha = \frac{1}{2w_0^{2-\alpha}}\left(-\frac{\partial^2}{\partial x^2}-\frac{\partial^2}{\partial y^2}\right)^{\alpha/2} + \frac{n_2 k}{n_0}|u|^2.$$

where

$$\hat{u}(k_x,k_y,z) = \iint dxdy \exp\left(-ik_x x - ik_y y\right) u(x,y,z)$$

We aim to solve Eq. (1) with the input in the form of the CAGVBs written in terms of the polar coordinates, $(r,\phi)$:

$$u(r,\varphi,z=0) = A_0 Ai\left(\pm\frac{r_0-r}{bw}\right)\exp\left(\pm d\frac{r_0-r}{bw}\right)\exp\left[-\frac{(r_0-r)^2}{w^2}\right]\left(\frac{r^m}{w^m}e^{im\varphi}\right), \quad (2)$$

where $A_0$ is the constant amplitude of the electric field, $Ai(\cdot)$ corresponds to the Airy function, $r$ is the radial coordinate, $r_0$ represents the radius of the primary Airy ring, and $w$ is a scaling factor, $b$ is the distribution factor parameter that makes the CAGVBs tend to ring Airy vortex beams when it is a low value; $0 \leq d < 1$ is the exponential truncation factor which determines the propagation distance, $\varphi = \arctan(y/x)$ is a spiral phase and $m$ is the topological charge of the optical vortex. The $\pm$ arrangement corresponds to inward and outward acceleration of the CAGVBs (in general, the CAGVBs in this study are inward acceleration cases). We cannot get the $u(r,\varphi,z)$ analytically with the initial electric fields $u(r,\varphi,z=0)$ in the FNSE optical system. Fortunately, numerical simulations are carried out to perform the propagation numerical calculation and show the simulation findings of the CAGVBs in the FNSE optical system. By taking Eq. (2) as an initial electric field, we solve Eq. (1) by using the fast Fourier transform method and obtain the numerical electric field $u(r,\varphi,z)$ of the propagation. In the simulations, we assume that

$\alpha = 1.5$, $m = 1$, $n_0 = 1.45$, $b = 0.1$, $\lambda = 532 \times 10^{-6}$ mm, $d = 0.1$, $r_0 = 1$ mm, $w = 1$ mm. The Rayleigh distance is $Z_R = kw^2/2$. The coefficient for the Kerr nonlinearity $n_2 = 2.6 \times 10^{-16}$ cm$^2$W$^{-1}$ leads to a critical power of a Gaussian beam for self-focusing $P_{cr} = \dfrac{3.77\lambda^2}{8\pi n_0 n_2}$. Using these parameters, the CAGVBs can be autofocusing at a certain position, and show some interesting propagation properties.

## 3. On-axis CAGVBs propagating in the FNSE optical system

Figure 1 depicts the numerical results of the CAGVBs with on-axis vortex propagating in the FNSE optical system with an input power $P_{in} = 0.06P_{cr}$. As shown in Fig. 1(a) and Figs. 1(b1)-1(b4), the lateral acceleration of the CAGVBs is attained and the energy rushes in an accelerated fashion toward the focus. It is quite clear that the CAGVBs sharply autofocus after a certain distance of propagation and the focal point (we can see that the CAGVBs are focused at a certain position, which is called the focal point) is located at about the distance z=4.8$Z_R$ and the light intensity increases to the maximum $4.88 \times 10^7$ cd. After the focusing position, it shows that the center of the transverse strength profile is always hollow, and the vortex forms a constant hollow channel along the propagation axis. When the CAGVBs propagate in the FNSE optical system, the process of the focus formation can be described by the action of mainly nonlinear physical effects: the nonlinear optical effect acts against on diffraction and makes the beam focus on itself. The intensity of the beam is usually the highest on the axis, located at the center of the beam. The curvature effect of the wave front is similar to that of a lens, with the difference that the effect is cumulative and can cause the beam to focus automatically. Figure 2 shows the three-dimensional plot of the surface corresponding to equal values of the intensity of the CAGVBs along the propagating axis in the FNSE optical system and the intensity distribution of the CAGVBs versus the propagation distance.

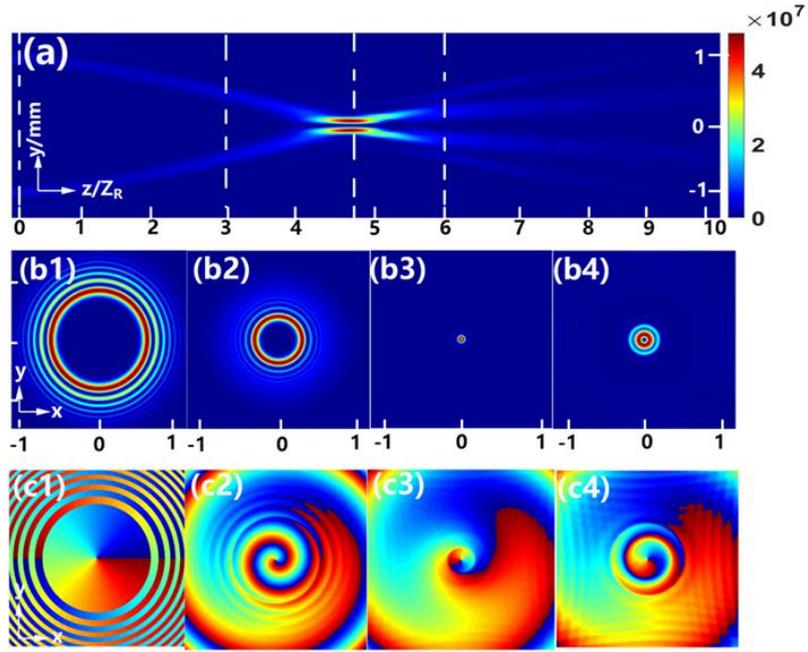

Fig. 1. (color online) Numerical demonstrations of the CAGVBs propagating in the FNSE optical system with $b=0.1,\ m=1,\ \alpha=1.6,\ P_{in}=0.06P_{cr}$.

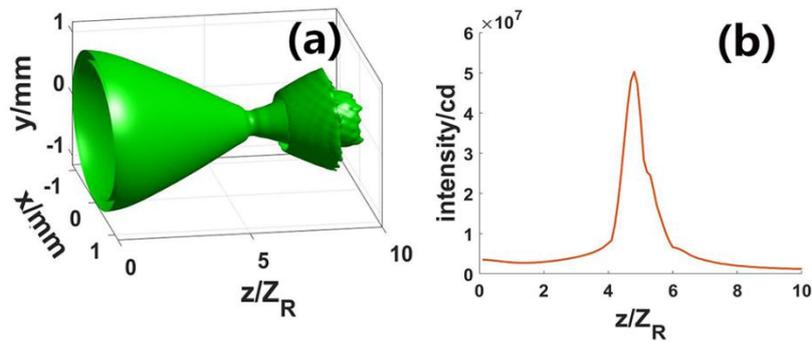

Fig. 2. (color online) (a) The corresponding iso-intensity distribution of the CAGVBs. (b) the intensity distribution of the CAGVBs versus the propagation distance. All parameters are the same as those in Fig. 1.

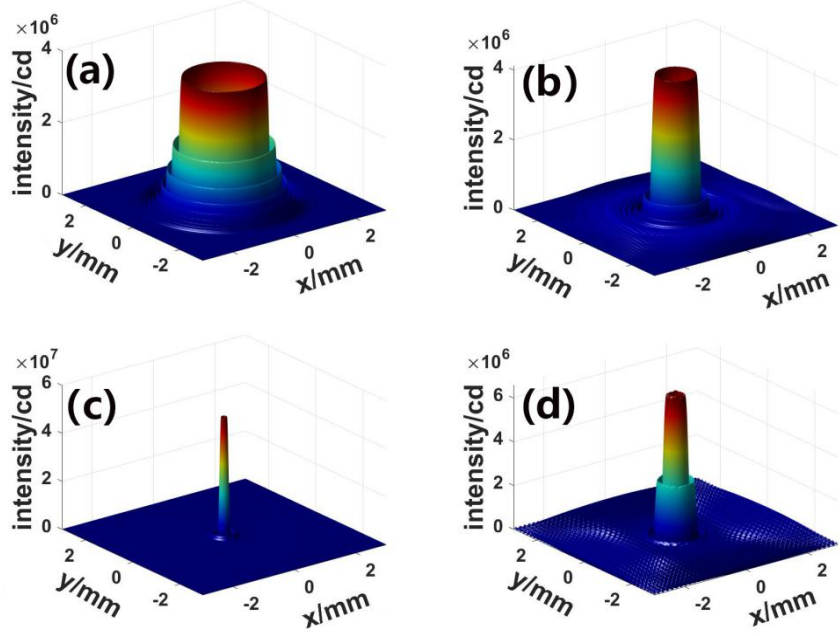

Fig. 3. (color online) The transverse intensity profiles for the CAGVBs at different propagation distances with (a)-(d) $0Z_R$, $3Z_R$, $4.8Z_R$, and $6Z_R$, respectively. All parameters are the same as those in Fig. 1.

The profiles for the case of different propagation distances with $b=0.1$, $m=1$, $P_{in}=0.06P_{cr}$ are shown in Fig. 3. In Figs. 3(a)-3(d), the intensity distributions at the positions $0Z_R$, $3Z_R$, $4.8Z_R$, and $6Z_R$ demonstrate the CAGVBs firstly undergoing autofocusing after a short distance, and the intensity reaches a sufficiently high value to trigger a focal point in the FNSE optical system. When the diffraction effect in a medium is canceled by the Kerr nonlinearity, the cross section undergoes compression during the main rings of the focusing. In Fig. 3(c), the main ring is compressed to the minimum in the presence of the nonlinear effect and the intensity of the CAGVBs is about more than 14 times that of the initial one. After autofocusing occurs, the number of outer rings increases and the radius of the main ring expands shown in Fig. 3(d).

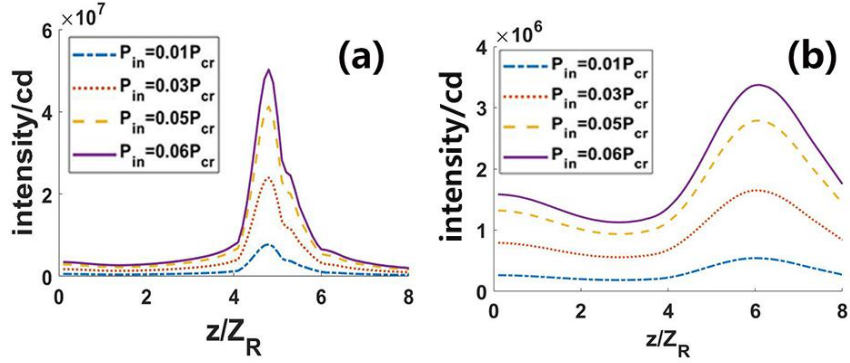

Fig. 4. (color online) The intensities of the CAGVBs as a function of the distribution factor $b$ with different initial powers, (a) $b = 0.1$ and (b) $b = 0.3$. The other parameters are $m = 1$ and $\alpha = 1.6$.

For fixing Lévy index $\alpha$, the distribution factor $b$ can greatly affect the autofocusing properties of the CAGVBs. It is different from ring Airy Gaussian beams with a spiral phase in the Kerr medium[29], the intensity of the CAGVBs always increases with a smaller distribution factor while decreases first with a bigger distribution factor. For $b = 0.1$, the intensities as functions of the propagation distance with different initial powers are described in detail in Fig. 4(a). As the power ratio is increased, the distribution of the intensity along the beam propagation changes significantly. For example, at $P_{in} = 0.01 P_{cr}$, the peak intensity is about $7.8 \times 10^6$ cd while the peak intensity is about $4.88 \times 10^7$ cd when $P_{in} = 0.06 P_{cr}$ in the same propagation distance $z = 4.8 Z_R$. It is shown that as the power grows continuously, the peak intensity gradually increases due to the nonlinear effect. Figure 4(b) demonstrates the intensities with diverse initial powers for $b = 0.3$. The properties are similar to what is found in Fig. 4(a), that is to say the CAGVBs have a higher peak intensity than that for the higher power ratio with the same distribution factor $b$. Moreover, it is also shown that the peak intensity decreases and the focal length increases greatly as increase of the distribution factor $b$ no matter what the value of the power is.

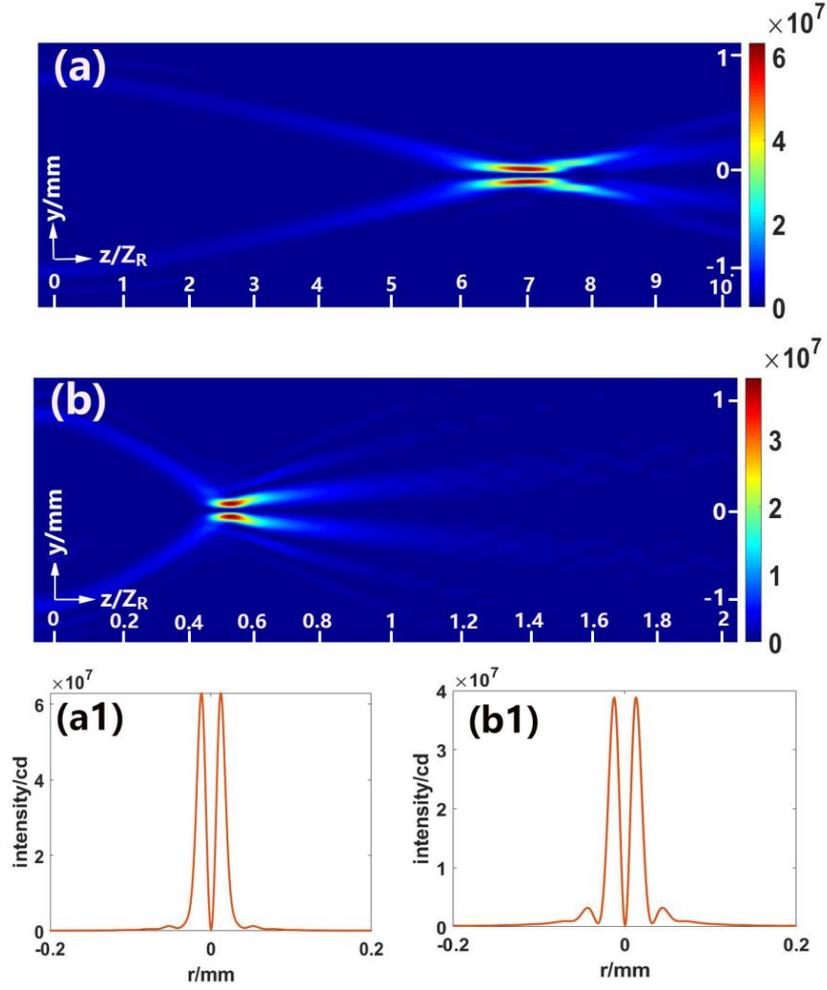

Fig. 5. (color online) (a) and (b) numerically simulated side-view propagation of the CAGVBs propagation, (a1) and (b1) show the intensities of the focal plane, respectively, for $\alpha = 1.5$ and $\alpha = 1.8$. The other parameters are $m = 1$, $b = 0.1$, and $P_{in} = 0.06 P_{cr}$.

Next, we perform some numerical simulations for the CAGVBs in the FNSE optical system. Figures 5(a) and 5(b) demonstrate the numerically simulated side-view propagation of the CAGVBs for different Lévy indexes. Figures 5(a1) and 5(b1), respectively, show the intensities of the focal plane for $\alpha = 1.5$ and $\alpha = 1.8$. Before autofocusing occurs, the parabolic trajectory is distorted by comparing Fig. 5(a) with Fig. 5(b). After autofocusing occurs, Figs. 5(a) and 5(b) show that the intensities of the profile keep the dark hollow rings, and dark hollow channels form because of the spiral phase on the axis along the propagation. Moreover, for $\alpha = 1.5$ and $\alpha = 1.8$, the CAGVBs autofocus sharply at a certain distance, and then diffract slightly. The diameter of the hollow channel for $\alpha = 1.8$ is bigger and diffracts faster than that of the case for $\alpha = 1.5.$ Figures 5(a1) and 5(b1) show the rings of the CAGVBs for

$\alpha = 1.8$ are bigger than those of the case $\alpha = 1.5$ and the intensities of the focal plane are greater than those of the case $\alpha = 1.8.$ We can find that with the increase of the Lévy index, the focal intensity decreases while the focal length increases by comparing Fig. 5(a1) with Fig.5(b1).

Figures 6(a) and 6(b) present the focal intensities and the focal lengths under different Lévy indexes, respectively. Figure 6(a) shows the focal intensity is about $1.73 \times 10^8 \, \text{cd}$ at about $\alpha = 1.4.$ Then the focal intensity rapidly decreases with the increase of the Lévy index, and the focal intensity is about $6.5 \times 10^6 \, \text{cd}$ at about $\alpha = 2.$ More important is that the focal intensity for the case of $\alpha < 2$ is always greater than that for $\alpha = 2.$ This means the focal intensity of the CAGVBs in a FNSE optical system is always larger than that in a standard nonlinear Schrödinger equation. On the other hand, with the increase of the Lévy index, the CAGVBs can propagate to the shorter distance before focusing, contributing to a substantial reduction of the focal length, which is shown in Fig. 6(b).

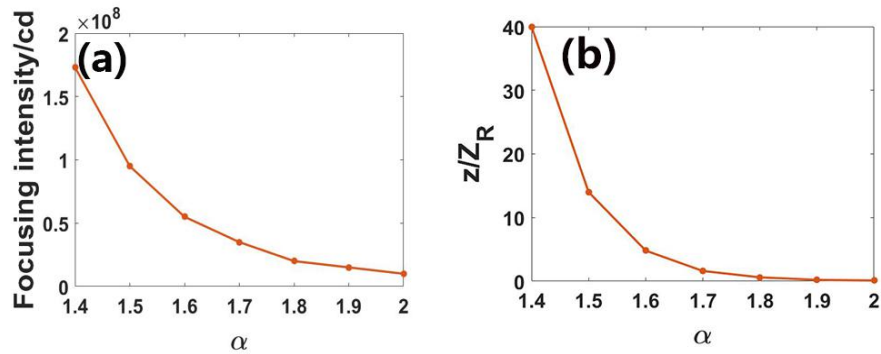

Fig. 6. (color online) (a) and (b) are the focal intensity and the focal length under different Lévy indexes, respectively. Other parameters are the same as those in Fig. 5.

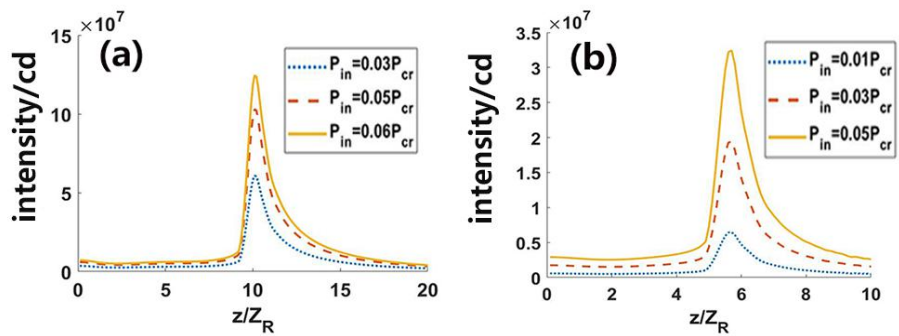

Fig. 7. (color online) The intensities of the CAGVBs as a function of the $z/Z_R$ with different

initial powers, (a) $\alpha = 1.5$ and (b) $\alpha = 1.9$. The other parameters are $m = 1$ and $b = 0.1$.

For $\alpha = 1.5$ the intensities as functions of the propagation distance with different initial powers are detailed in Fig. 7(a). It is shown that as the power is increased continuously, the peak intensity gradually increases due to the nonlinear effect. The CAGVBs have higher peak intensity than that for the higher power ratio with the same Lévy index. Figure 7(b) demonstrates the intensities with diverse initial powers for $\alpha = 1.9$. We can see that with the increase of the Lévy index, the focal intensity attenuates and the focal length greatly decreases. These observations are in good agreement with the numerical simulations of the relationship between the focal intensity and the focal length under different Lévy indexes shown in Fig. 6.

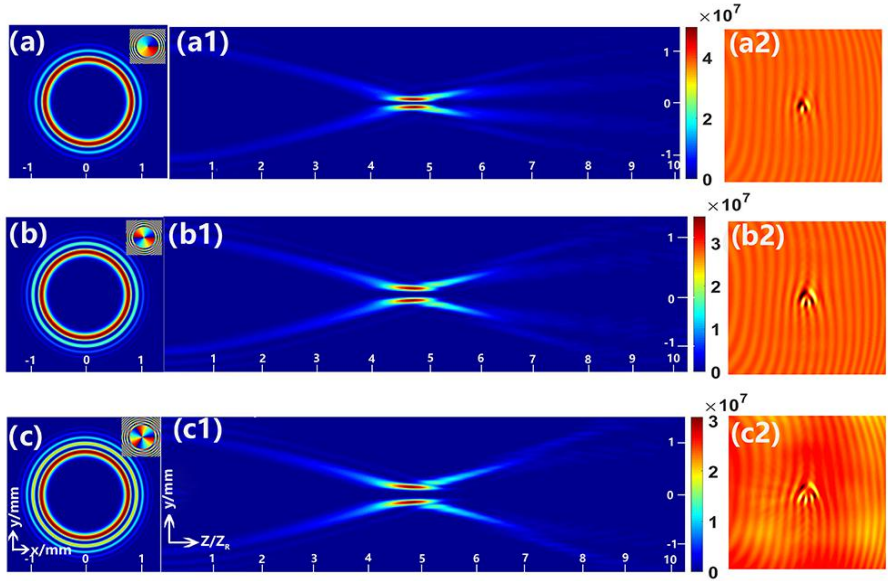

Fig. 8. (color online) Numerical demonstrations of the CAGVBs propagating with $b = 0.1$ versus different topological charges. (a)-(c) the intensities of the initial input CAGVBs, (a1)-(c1) numerically simulated side-view propagation distributions of the CAGVBs, (a2)-(c2) interference intensities of the CAGVBs in the focal plane for m=1, 2, 3, respectively. The other parameters are $\alpha = 1.6$ and $P_{in} = 0.06 P_{cr}$.

Next we find that the number of the topological charge of the CAGVBs affects the propagating properties in the FNSE optical system as well. The initial intensities and phases are shown in Figs. 8(a)-8(c), in which the CAGVBs carry the OAM with $m\hbar$ at the initial plane. With bigger topological charges, the OAM is larger, and the intensity of the secondary ring becomes greater as well. When the CAGVBs propagate in the FNSE optical system, the autofocused CAGVBs remain in the dark hollow channel on the axis, and the hollow region in the center persists even at the focal point or after the focal point. By increasing the number of the topological charge for the CAGVBs in the FNSE optical system, we can find that the focal point of the

CAGVBs is slightly shifted towards the laser source while the intensity of the CAGVBs decreases distinctly in Figs. 8(a1)-8(c1). On the other hand, we can find that the diameter of the hollow channel becomes larger with the increase of the topological charge. With the interference intensities of the CAGVBs in the focal plane (see Figs. 8(a2)-8(c2)), we can see that the OAM can be still preserved along the propagation, which shows the vortex can stably propagate.

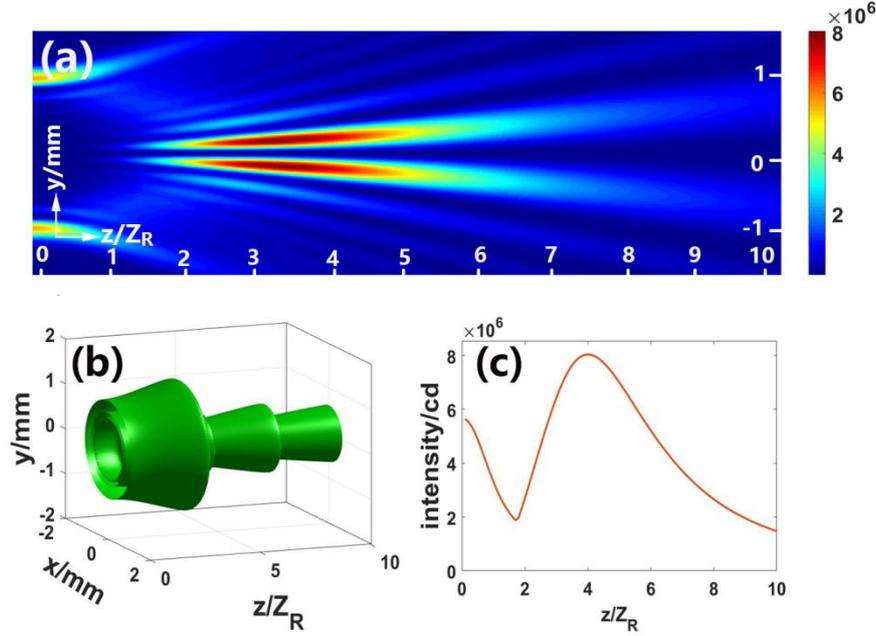

Fig. 9. (color online) (a) Numerical demonstrations of the CAGVBs with outward acceleration propagating in self-defocusing nonlinearity. (b) The corresponding iso-intensity distribution of the CAGVBs corresponding to (a). (c) the intensity distribution of the CAGVBs versus the propagation distance. Other parameters are the same as those in Fig. 1.

Compared with the inward acceleration in Fig. 1, the CAGVBs can accelerate outward by replacing focusing with self-defocusing effect in system, whose outcomes are given in Fig. 9. It is shown that the CAGVBs bend outward and experience autodefocusing. Due to the self-defocusing effect, CAGVBs with outward acceleration focus faster than that of inward acceleration (Fig. 9(a)). However, it is different from RAiG beams with outward acceleration propagating in the Kerr medium[29], they focus slower than the inward acceleration CAGVBs, and the maximum peak intensity is lower as well. Figure 9(b) shows the three-dimensional plot of the surface corresponding to equal values of the intensity of the CAGVBs with outward acceleration along the propagating axis in the FNSE optical system and Fig. 9(c) presents the intensity distribution of the CAGVBs versus the propagation distance.

## 4. Off-axis propagation dynamics of autofocusing CAGVBs with positive vortex pairs in the FNSE optical system

Using the split-step Fourier method, we assume that the vortex pairs of off-axis CAGVBs distribute symmetrically about the origin in the x-axis, and the phase $\phi_k = 0$. Here, we consider the fields in the initial plane can be expressed as:

$$u(r,\varphi,z=0) = A_0 A_i\left(\frac{r_0-r}{bw}\right)\exp\left(d\frac{(r_0-r)}{bw}\right)\exp\left[-\frac{(r_0-r)^2}{w^2}\right]\left[\frac{(re^{i\phi_k}+r_k)(re^{i\phi_k}-r_k)}{w^2}\right], \quad (3)$$

where ($r_k, \varphi_k$) denotes the location of the optical vortex. In the simulation, we assume that the positive vortex pairs are off-axis with $r_k = 0.6mm$, an input power $P_{in} = 0.06P_{cr}$ and the other parameters are the same as those in Fig. 1.

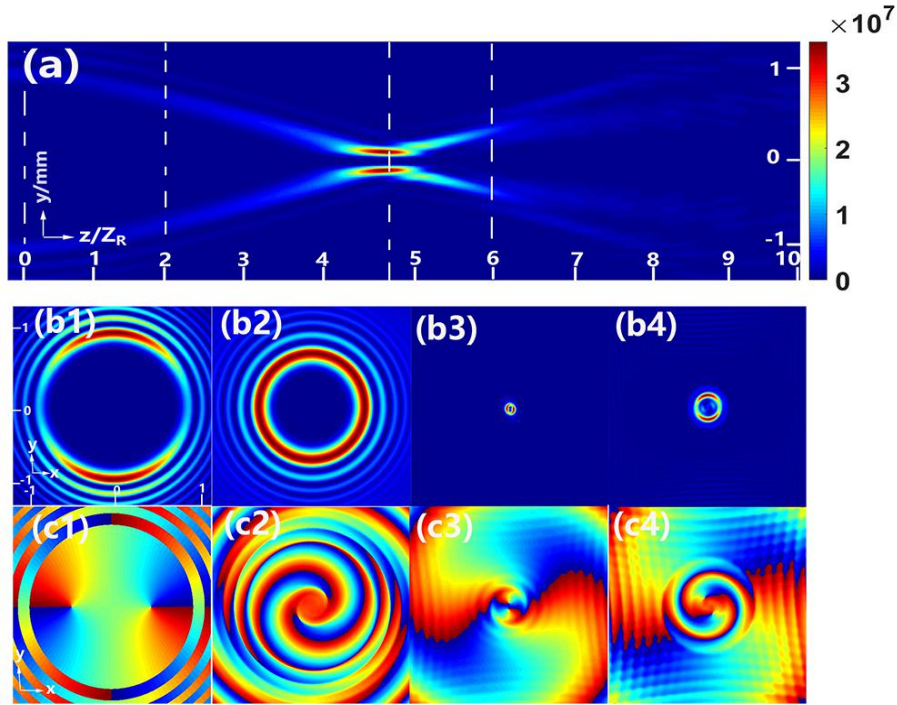

Fig. 10. (color online) All are the same as those in Fig. 1 except off-axis positive vortex pairs and the input power.

Figure 10 shows the off-axis propagation of the CAGVBs with positive vortex pairs in the FNSE optical system by inputting the solution of Eq. (3). From the side-view propagation (Fig. 10(a)), we can see that the CAGVBs also sharply autofocus after a certain propagation distance. The focal point is located at about the distance $Z = 4.7Z_R$ and the light intensity increases to the maximum about $4.15 \times 10^7$ cd. After autofocusing occurs, the autofocused beams remain the dark hollow channel on the axis, and the hollow region in the center persists even at the focal point or after the focal point, while the distance between the vortex pairs increases gradually and the intensity of the light spot becomes weaker. In Figs. 10(b1)-10(b4), the off-axis optical vortex pairs are rotated relative to the propagation axis. Interestingly, they are also forced to move to near the center and the intensity of the light spot increases accordingly. The intensity pattern of the off-axis propagation of the CAGVBs with positive vortex pairs is an oval-shaped light spot due to the influence of the off-axis vortex at the focal plane. In Figs. 10(c1)-10(c4), the evolution of off-axis phase patterns always moves in a counterclockwise screw on-axis in the FNSE optical system.

## 5. Conclusions

In summary, we have presented the propagation dynamics of the CAGVBs in the FNSE optical system by using the split-step Fourier method. By combining fractional diffraction with nonlinear effects, the influences of the Lévy index, the distribution factor, the power ratio and the number of the topological charge on propagation dynamics of the CAGVBs are investigated. Numerical simulations indicate that the autofocusing effect becomes weaker when the Lévy index increases. With the increase of the Lévy index, the focal intensity attenuates and the focal length greatly decreases. It should be noted that the focal intensity of the CAGVBs in a FNSE is always greater than that in a standard NLSE by adjusting the Lévy index. Moreover, it has been found that the autofocusing effect becomes weaker when the distribution factor increases, and the CAGVBs have a higher peak intensity than that for the higher the power ratio with the same distribution factor. Then, one can also modify the autofocusing properties of the CAGVBs by appropriately selecting the number of the topological charge. Next, we also create the CAGVBs with outward acceleration, and find that the CAGVBs bend outward and show the autodefocusing properties. Finally, we show that the off-axis propagation dynamics of the CAGVBs with positive vortex pairs in the FNSE optical system. Our results can be used in many applications such as medical treatment, particle processing and optical communication.

**FUNDING INFORMATION**

This work was supported by the National Natural Science Foundation of China (11775083, 11374108, 61675001, 11947103, 12004081). Science and Technology Program of Guangzhou (No. 2019050001).

**CONFLICT OF INTEREST**

We declare that we have no conflict of interest.